\RequirePackage{lineno}

\documentclass[aps,pre,amsmath,amsfonts,amssymb,onecolumn]{revtex4}
\usepackage{bm}        

\usepackage[latin1]{inputenc}
\usepackage{graphicx}
\usepackage[english]{babel}
\usepackage[usenames,dvipsnames]{color}

\usepackage{amsfonts}
\usepackage{amsmath}
\usepackage{amssymb}

\usepackage{color}

\usepackage{soul}

\newcommand{\beq}{\begin{equation}}
\newcommand{\eeq}{\end{equation}}
\newcommand{\nn}{\nonumber}

\newcommand{\ra}{\rightarrow}

\makeatletter
\@ifundefined{textcolor}{}
{%
 \definecolor{BLACK}{gray}{0}
 \definecolor{WHITE}{gray}{1}
 \definecolor{RED}{rgb}{1,0,0}
 \definecolor{GREEN}{rgb}{0,1,0}
 \definecolor{BLUE}{rgb}{0,0,1}
 \definecolor{CYAN}{cmyk}{1,0,0,0}
 \definecolor{MAGENTA}{cmyk}{0,1,0,0}
 \definecolor{YELLOW}{cmyk}{0,0,1,0}
 }

\begin{document}


\title{Thermoforesis from generalized Caldeira-Leggett
models}

\author{Daniel Valente
$^{1,2}$
}
\email{valente.daniel@gmail.com}

\author{Maur\'icio Matos
$^{2}$
}

\author{Thiago Werlang
$^{2}$
}

\affiliation{
$^{1}$ 
Centro Brasileiro de Pesquisas F\'isicas, Rio de Janeiro, RJ, Brazil
}

\affiliation{
$^{2}$ 
Instituto de F\'isica, Universidade Federal de Mato Grosso, Cuiab\'a, MT, Brazil
}

\begin{abstract}
The standard Caldeira-Leggett model addresses the problem of Brownian motion in a thermal equilibrium environment. 
Here, we look for generalizations of the Caldeira-Leggett model to account for thermal gradients in the environment.
We devise two types of models, and discuss the advantages and limitations of each one. 
From both models, we find signatures of thermophoresis, i.e., particle transport due to a thermal gradient. 
In principle, our models can be employed to describe thermophoresis in quantum Brownian particles, an open problem so far.
\end{abstract}


\maketitle

\section{Introduction}\label{sec:introduction}

The standard Caldeira-Leggett model (sCLm) has been successfully designed to describe the presence of dissipation in the quantum tunneling of a macroscopic variable such as magnetic flux or current in superconducting devices \cite{bookcaldeira}.
In recent years, these superconducting devices have matured to become some of the most widespread hardware for quantum computation \cite{nobel2025}.
From our theoretical point of view, nevertheless, the most important aspect is the versatility of the model.
To begin with, it encompasses ohmic superconducting circuits and charged particles in metallic environments \cite{hedegaard1987quantum,guinea,weiss2012quantum}, as well as superohmic Abraham-Lorentz forces \cite{barone}.
Additionally, extensions of the sCLm have also been devised to include not only pairs of quantum Brownian particles \cite{duarte2006effective,duarte2009effective,valente2010thermal}, but also to address fermionic baths \cite{hedegaard1987quantum}.
The collision model, an alternative form of system-plus-reservoir approach which does not assume harmonic oscillators as a model for the environment, has also been investigated to describe the quantum dissipative dynamics of solitons, such as domain walls or vortices in magnetic or superconducting media \cite{castroneto92, castroneto931,castroneto932, castroneto95}.

Here, we generalize the Caldeira-Leggett model with the goal of describing the effects of a thermal gradient on the Brownian motion of the particle.
We are particularly interested in describing thermophoresis, where a thermophobic force pushes the particle to the colder side of an environment subjected to a thermal gradient.
This kind of phenomenon, while still under investigation in the classical regime, has only recently been extended to the quantum regime \cite{pre26}.
In the context of quantum thermodynamics, the control of heat flow in the presence of quantum effects has received attention, especially with superconducting circuits \cite{pekkola}.
What Ref.\cite{pre26} shows is that thermal gradients not only affect the flow of heat through quantum systems, but also the flow of matter and information in both real and Hilbert spaces.
The quantum thermophoresis shown in Ref.\cite{pre26} leads to dissipative self-organization, thus opening a path towards the processing of quantum information based on heat flows, as suggested by the concept of thermodynamic computing \cite{thermocomp1,thermocomp2,thermocomp3}.

The generalized models we discuss here are classical.
Nonetheless, they allow in principle a quantum treatment.
This is important because the thermophoretic behavior of a quantum Brownian particle is still an open problem.
The quantum master equations used in Ref.\cite{pre26} only apply to quantum systems with a discrete energy spectrum, and a finite number of thermal baths acting on specific regions of the particle's surroundings.
Our models, on the contrary, describe continuously varying temperatures in space and can be employed to a quasi-free particle (i.e., a system with continuous energy spectrum).
In other words, our models open the path for the study of thermophoresis in quantum Brownian particles.

The paper is organized as follows.
In Sec. 2, we revisit key signatures of thermophoresis, as they show up in Langevin and Fokker-Planck dynamics.
In Sec. 3, we present our simplest model, and discuss its consequences and limitations.
In Sec. 4, we present and analyze our more sophisticated model.
Finally, in Sec. 5 we summarize our conclusions and perspectives for future studies.

\section{Signatures of thermophoresis}

\subsection{The Langevin force}
\label{sec:2.1}

The average Langevin force becomes finite, and becomes a measure of the thermophoretic force.
To see that, we write the Langevin force $f(x,t)$ as a product involving the space-dependent diffusion coefficient $D(x)$ and the fluctuating time-dependent signal $\xi(t)$,
\begin{equation}
    \langle f(x,t) \rangle = \langle \sqrt{2D(x)} \xi(t) \rangle \neq \langle \sqrt{2D(x)}\rangle \langle \xi(t) \rangle.
\end{equation}
The inequality is due to $x=x(t)$ being correlated with the bath fluctuations, from time $0$ to time $t$.
This evidences a non-vanishing average force, even though the fluctuating signal has zero average, $\langle \xi(t) \rangle=0$.

Let us now discuss a simple model for this average.
As presented in Ref.\cite{pre26},
we can split the Langevin force as a contribution coming from the left side ($f_L$) and another coming from the right side ($f_R$) of the Brownian particle, so that
\beq
f(x,t) = f_L - f_R = \sqrt{T(x_L)} \xi_L(t) - \sqrt{T(x_R)} \xi_R(t).
\label{eqfsplitLR}
\eeq
Here, $T(x)$ is the temperature, and 
$\xi_{L,R}(t)$ satisfy 
$\langle \xi_{k}(t) \xi_{k'}(t')\rangle = 2\eta k_B \delta(t-t') \delta_{k,k'}$ for $k=R,L$
\cite{matsuo2000}.
The damping rate is $\eta$ and $k_B$ is the Boltzmann's constant.
Without loss of generality, we can decompose the full Brownian fluctuating signal (having zero average) as $\xi(t) = \xi_L(t) - \xi_R(t)$, where $\xi_{L,R}(t)$ each describe nonnegative fluctuating signals.
Because $\langle \xi(t) \rangle = 0$, we have that 
$\langle \xi_L(t) \rangle = \langle \xi_R(t) \rangle$.
If the particle has a linear size $2r$, then $x_{L,R} = x\mp r$.
Taylor expansion to first order in $r$ gives us
\beq
f(x,t) = \sqrt{T(x)}(\xi_L(t) - \xi_R(t)) 
- \frac{T'(x)}{\sqrt{T(x)}} \frac{r}{2} (\xi_L(t) + \xi_R(t)).
\eeq
The thermophoretic force appears as
$\langle f(x,t) \rangle = - (T'(x)/\sqrt{T(x)}) r \langle\xi_L(t)\rangle$.
But 
$\langle \xi_L(t)\rangle = \langle f_L \rangle /\sqrt{T(x)} = p A/\sqrt{T(x)}$,
where $p$ is the local pressure and $A$ is the cross-section of the Brownian particle.
Within such a simplified model, one finds that the average thermophoretic force acting on a classical Brownian particle reads
\beq
\langle f(x,t) \rangle = - \frac{T'(x)}{T(x)} \frac{pV}{2},
\label{fLclass}
\eeq
where $V=2rA$ is the volume of the particle.
For a classical point-like particle, $V\ra 0$, thermophoresis cannot take place.
The negative sign, $-T'(x)$, results in a force that pushes towards the cold side of the nonequilibrium bath.

\subsection{The Fokker-Planck equation}
\label{sec2.2}

In the overdamped regime, $M\ddot{x} \ll \eta\dot{x}$, the Langevin equation becomes
\beq
\dot{x} = -\frac{V'(x)}{\eta}+
\sqrt{\frac{2D(x)}{\eta^2}} \xi(t),
\eeq
where 
$V(x)$
is an external potential,
and
$\langle \xi(t) \xi(t')\rangle = \delta(t-t')$.
From the overdamped Langevin equation for the particle position, one can derive a Fokker-Planck equation for the probability density $P(x,t)$,
\begin{equation}
\frac{\partial P(x,t)}{\partial t}
=
\frac{\partial}{\partial x}
\left[
\frac{V'(x)}{\eta}P(x,t)
+
\frac{\partial}{\partial x}
\left[
\frac{D(x)}{\eta^2} 
P(x,t)
\right]
\right].
\end{equation}

The derivative of the temperature field appears due to
\beq
D'(x) = 
\eta k_B T'(x).
\eeq
With the help of the continuity equation,
$\partial_t P = -\partial_x J$,
where
$J=-\eta^{-1} V'(x)P-\eta^{-2} D(x)\partial_x P- \eta^{-2} D_T P \partial_x T(x)$,
and
$D_T = \eta k_B$,
we obtain for the steady state, $J=0$, of a free particle, $V'(x)=0$, that
\beq
P'_{ss}(x) = - S_T(x) T'(x) P_{ss}(x),
\label{eqsoret}
\eeq
where
$S_T(x) = D_T/D(x) = 1/T(x)$ is the so called Soret coefficient.

\section{Generalized Caldeira-Leggett model I}

Our generalized Caldeira-Leggett model I (gCLm-I) is strongly based on intuition derived from the model discussed in Eqs.(\ref{eqfsplitLR})-(\ref{fLclass}).
As discussed in the previous section, an asymmetric Langevin force arises from the fact that the hot region on the surroundings of the particle knocks it more strongly than the cold region, on average.
To mimic that, we assume that the oscillators in the bath, which can be seen as the Fourier components of the fluctuating Langevin force, are constantly being pushed by an external force.
We can interpret this external agent pushing the bath oscillators as the system which provides the energy source for the (externally imposed) thermal gradient.
This external force depends on the local temperature gradient at the position of the system (the Brownian particle).

\subsection{The Langevin equation for the gCLm-I}

Let us first recall the standard Caldeira-Leggett model Hamiltonian (sCLm) \cite{bookcaldeira},
\begin{equation}
    H_{sCLm} = \frac{p^2}{2M} + V(x) + \sum_k \left[ \frac{p_k^2}{2 m_k} + \frac{1}{2}m_k \omega_k^2 \left(q_k - \frac{c_k}{m_k \omega_k^2 }x\right)^2\right],
\end{equation}
where $x$ and $p$ are the position and momentum of the system with mass $M$, and the oscillators have mass $m_k$, frequencies $\omega_k$, positions $q_k$ and momenta $p_k$.
The standard equations of motion read
\begin{equation}
    M\ddot{x} = -V'(x) + \sum_k c_k q_k - \sum_k \frac{c_k^2}{m_k \omega_k^2} x,
    \label{eqsys}
\end{equation}
for the system, and
\begin{equation}
    m_k\ddot{q}_k = -m_k \omega_k^2 q_k + c_k x,
\end{equation}
for the bath oscillators.

Our gCLm-I consists in assuming that an external agent is constantly pushing the bath of oscillators with a force
\begin{equation}
    f^{(k)}_{ext} = - \alpha_k T'(x),
\end{equation}
so that the modified equations of motion for the bath oscillators are
\begin{equation}
   m_k\ddot{q}_k = -m_k \omega_k^2 q_k + c_k x - \alpha_k T'(x). 
\label{qkfext}
\end{equation}
The external coupling constants $\alpha_k$ have units of energy per temperature, just as $k_B$ and $pV/T$ [in Eq.(\ref{fLclass})].
In the limit $\alpha_k \ra 0$, we reobtain the sCLm.

Note that we started from the external force, $f^{(k)}_{ext}$, rather than an external Hamiltonian, $H_{ext}$.
This is to keep the equation of motion for the system unaltered, since the heat sources push the bath out of equilibrium but not directly the system itself.
If we want to conceive a Hamiltonian that gives rise to $f^{(k)}_{ext}$, while preserving the system equation of motion, we shall restrict the model to the particular case of a constant gradient,
$T''(x) = 0$,
or
$T'(x) = T'(x_0) = T'_0$.
In that case, we can assume that
$H_{ext} = \sum_k \alpha_k T'_0 q_k$.
As a consequence, the present model may only be quantized for the special case of a constant gradient.
This makes sense to us: if the temperature derivative depends on the system position, it means that we are considering an external agent that has full information on the system's position $x$.
This is allowed in classical mechanics, but not in quantum mechanics.

The solution of Eq.(\ref{qkfext}) is
\begin{equation}
    q_k(t) = q_k^{(0)}(t)
    +
    \frac{1}{m_k\omega_k}
\int_0^t ds
\sin[\omega_k(t-s)][c_k x(s) - \alpha_k T'(x(s))],
\label{solqk}
\end{equation}
where $q_k^{(0)}(t) = q_k(0)\cos(\omega_k t) + (p_k(0)/m_k\omega_k) \sin(\omega_k t)$ is the homogeneous solution.

By substituting Eq.(\ref{solqk}) in (\ref{eqsys}), we find
\begin{equation}
    M\ddot{x}+V'(x)=F_0 + F_{fb}+F_{CT},
\end{equation}
where $F_0 = \sum_k c_k q_k^{(0)}(t)$ is related with the standard Langevin force (as we will discuss further below), $F_{CT} = - \sum_k \frac{c_k^2}{m_k \omega_k^2}x$ is the counter-term force that will cancel out soon, and $F_{fb}$ is a feedback force,
\begin{equation}
    F_{fb} = \sum_k \frac{c_k}{m_k \omega_k} \int_0^t ds \sin[\omega_k(t-s)][c_k x(s) - \alpha_k T'(x(s))].
\end{equation}
By using that 
\beq
\int_0^t ds \sin[\omega_k(t-s)] f(s) = \frac{f(t)}{\omega_k}-\frac{1}{\omega_k}\frac{d}{dt} \int_0^t ds \cos[\omega_k(t-s)]f(s), 
\label{eqtruqueseno}
\eeq
we get that
\begin{equation}
    F_{fb} = -F_{CT}+F_{th} + F_{mem}.
\end{equation}
The counter-term force has finally canceled out.
The thermophoretic force reads
\begin{equation}
    F_{th} = - \sum_k \frac{c_k \alpha_k}{m_k\omega_k^2} T'(x).
\label{thI}
\end{equation}
And the memory force reads
\beq
F_{mem} = - \sum_k \frac{c_k}{m_k \omega_k^2}\frac{d}{dt}\int_0^t ds \cos[\omega_k(t-s)] [c_k x(s) - \alpha_k T'(x(s))].
\eeq
It gives rise to the dissipative force.
To see that, we make 
$\alpha_k = \tilde{\alpha} c_k$,
thus obtaining
\beq
F_{mem} = - \frac{d}{dt} \int_0^t ds \sum_k \frac{c_k^2}{m_k\omega_k^2} \cos[\omega_k(t-s)][x(s)-\tilde{\alpha} T'(x(s))].
\eeq
We can now make use of the spectral function,
\beq
J(\omega) = \frac{\pi}{2}\sum_k \frac{c_k^2}{m_k \omega_k} \delta(\omega-\omega_k).
\eeq
We define
\beq
K(t-s) \equiv \frac{2}{\pi}\int_0^\infty d\omega \frac{J(\omega)}{\omega} \cos[\omega(t-s)],
\label{defK}
\eeq
so that
\beq
F_{mem} = -\frac{d}{dt}\int_0^t ds K(t-s) [x(s)-\tilde{\alpha} T'(x(s))].
\label{eqmem}
\eeq
We choose an ohmic spectral function, 
$J(\omega) = \eta \omega$, 
and get that 
$K(\tau) = 2\eta \delta(\tau)$.
Also, we use that
$\frac{d}{dt}\int_0^t ds \delta(t-s)f(s) = \dot{f}(t)/2 + f(0)\delta(t)$, 
to obtain
\beq
F_{mem} = F_{diss}+\tilde{F}_{0}.
\eeq
The dissipative force is
\beq
F_{diss}=-\eta \frac{d}{dt}[x(t)-\tilde{\alpha}T'(x(t))] 
\equiv -\eta_{eff}[x]\ \dot{x},
\label{eq:etaeff}
\eeq
where the effective dissipation rate is
$\eta_{eff}[x] = \eta(1-\tilde{\alpha}T''(x))$.
The second term, 
$\tilde{F}_0 = -[x(0)-\tilde{\alpha}T'(x(0))]2\eta\delta(t)$, 
is added to $F_0$, yielding the Langevin force of the sCLm, namely,
$F_L = F_0 + \tilde{F}_0 = \sum_k c_k \left[ \tilde{q}_k(0)\cos(\omega_k t)+ (p_k(0)/m_k\omega_k)\sin(\omega_k t)\right]$,
with 
$\tilde{q}_k(0)\equiv q_k(0)-\frac{c_k}{m_k\omega_k^2}[x(0) - \tilde{\alpha}T'(x(0))]$ representing the displaced equilibrium position of the oscillators.

In summary, we find that
\begin{equation}
M\ddot{x}=-V'(x) -\eta_{eff}[x] \dot{x}-\kappa T'(x) + F_L(t).
\label{gCLMI}    
\end{equation}
Let us now discuss term $F_L(t)$, describing the standard Langevin force.
By assuming a thermal equilibrium distribution for the bath oscillators, $\langle \tilde{q}_k(0) \rangle = 0 = \langle p_k(0)\rangle$,
along with
$\langle p^2_k(0)\rangle/m_k = k_B T(x(0))$,
and
$m_k\omega_k^2\langle \tilde{q}^2_k(0)\rangle = k_B T(x(0))$,
we find that
$\langle F_L(t)\rangle = 0$,
and
$\langle F_L(t) F_L(t') \rangle = 2\eta k_B T_0 \delta(t-t')$,
where
$T_0 = T(x(0))$.
The fact that the correlation strength of the fluctuating force $F_L(t)$ depends on $T_0$, and not on $T(x(t))$, should be regarded as a zeroth-order approximation of the actual fluctuating force, i.e.,
$T(x) = T_0 + T'(x_0) (x-x_0) + (...) \approx T_0$.
Of course, this is a strong limitation of the present model we would like to avoid.
This issue is addressed in the following section.

We now go back to discuss the finite average (thermophoretic) force a little further, 
$F_{th} = -\kappa T'(x)$,
where
\beq
\kappa \equiv \sum_k \frac{c_k^2}{m_k \omega_k^2} \tilde{\alpha}.
\eeq
In order to guarantee the convergence for $\kappa$, we note that
$\sum_kc_k^2/m_k\omega_k^2 = \pi K(0)/2 = \pi \eta/\tau_R$, where $\tau_R \ra 0$ is the correlation time of the bath.
We thus have to impose
$\tilde{\alpha} = (\tau_R/\pi \eta) (pV/2T)$, if we want to get
$\kappa = pV/2T$,
as in the model we discussed in the previous section.
From this line of thought, we attribute to $\tilde{\alpha}$ a geometric meaning, i.e., it sets the effective volume of the Brownian particle.
The remaining terms in $\kappa$ depend on the bath of oscillators, which consist in the environment in the present model, phenomenologically described by the pressure and the local temperature. 
In the $\tau_R \ra 0$ limit, we also get that
$\tilde{\alpha}\ra 0$,
implying that
$\eta_{eff}[x] \ra \eta$.

Solving the non-linear Eq.(\ref{gCLMI}) for general $\tilde{\alpha}$ (in $\eta_{eff}[x]$) and $\kappa$ (in $F_{th}$), might give us interesting effects due to the dependence both on the first and the second derivatives of the temperature field.
Also, it might be worth investigating more general spectral functions, with finite $\tau_R$. 
Typically, one assumes that 
$J(\omega) = \eta_s \omega^s e^{-\omega/\omega_c}$,
where $\omega_c = \tau_R^{-1}$ is the cutoff frequency, and subohmic (superohmic) regimes are obtained with $s<1$ ($s>1$).
In those non-ohmic cases, memory effects will have an extra contribution due to the term $T'(x(s))$ in Eq.(\ref{eqmem}).

\subsection{The Fokker-Planck equation for the gCLm-I}

From the overdamped regime of Eq.(\ref{gCLMI}), we have that
\beq
\dot{x}=-\frac{V'(x)+\kappa T'(x)}{\eta_{eff}[x]} + \sqrt{\frac{2 D_0}{\eta_{eff}[x]}} \xi(t),
\eeq
where $D_0 = \eta k_B T_0$, and $\langle \xi(t) \xi(t')\rangle = \delta(t-t')$, along with $\langle \xi(t)\rangle = 0$.

Following the prescription from Sec.\ref{sec2.2}, we find that
\beq
\frac{\partial P_I(x,t)}{\partial t}
=
\frac{\partial}{\partial x}
\left[
\left(\frac{V'(x)+\kappa T'(x)}{\eta_{eff}[x]}\right) P_I(x,t)
+
\frac{\partial}{\partial x}
\left[
D_{eff}(x) 
P_I(x,t)
\right]
\right],
\eeq
where $D_{eff}(x) \equiv D_0/\eta^2_{eff}[x]$.
We notice that the spatial dependence of the temperature appears not only by means of $T'(x)$ (as a correction to the external potential), but also through $\eta_{eff}[x]$ (see Eq.(\ref{eq:etaeff})).
The space dependence of the effective diffusion coefficient, 
$D_{eff}(x)$
is now solely dependent on $\eta_{eff}[x]$, which in turn depends on the second derivative of the temperature.

At constant gradients, $T''(x) = 0$, we have that 
$\eta_{eff}[x]=\eta$, 
and a constant effective diffusion coefficient, 
$D_{eff}(x) = D_0/\eta^2 = k_B T_0/\eta$.
For a free particle, $V'(x)=0$, the steady state equation thus becomes
$P'_{ss,I}(x) = - (\kappa T'_0/k_B T_0) P_{ss,I}(x)$.
That is,
\beq
P_{ss,I}(x)=A^{-1} e^{-\frac{\kappa T'_0}{k_B T_0} x}, 
\eeq
where 
$A \equiv \int_{-L/2}^{L/2}du \exp[-(\kappa T'_0/k_B T_0) u] = (2k_B T_0/\kappa T'_0)\sinh(\kappa T_0' L/k_B T_0)$
is a normalization constant for the particle trapped in a region of size $L$ around $x_0=0$.
If $T_0'>0$ (resp. $T_0'<0$), the particle gets exponentially more concentrated at the cold region, $x<0$ (resp. $x>0$).

\section{Generalized Caldeira-Leggett model II}

The previous model, based on intuition derived from Eq.(\ref{eqfsplitLR}), assumes a thermal gradient introduced as a force acting on the bath of oscillators surrounding the Brownian particle.
However, the standard Caldeira-Leggett model (sCLm) uses the initial state statistics of the oscillators to introduce the temperature.
We now devise an alternative model, namely a generalized Caldeira-Leggett model II (gCLm-II), which addresses this issue.
Instead of a single driven bath of oscillators, we assume a {\it continuum} of thermal baths in a space of size $L$.
At each point $X$, a local temperature $T(X)$ is attributed to that local bath (i.e., to the statistics of the initial states of the oscillators of that local bath).

\subsection{The Langevin equation for the gCLm-II}

The generalized system-bath Hamiltonian in this case reads
\begin{equation}
H_{gCLm-II}=
\frac{p^2}{2M}+V(x)
+
\int \frac{dX}{L}\sum_k
\left[
\frac{p^2_k(X)}{2m_k}
+
\frac12 m_k\omega_k^2 Q^2_k(X,t)
\right],
\end{equation}
where
\begin{equation}
Q_k(X,t)=
q_k(X,t)-\frac{c_k}{m_k\omega_k^2}x(t)g(x(t)-X).
\end{equation}
Here, $L$ can be seen as the size of the universe where the Brownian particle lives (the eventual quantization volume), and $g(u)$ is a dimensionless weight function specifying how the surrounding baths act on the system, depending on its position $x(t)$.
If $g(u)\ra L\delta(u)$, the system becomes coupled only to the local bath at $X=x(t)$, as in the sCLm.
The full width at half maximum of the function $g(u)$ can be thought of as an effective volume of the Brownian particle, in terms of the spatially extended range of its sensing capacity of its own environment.

It turns out that defining
\begin{equation}
G(x,X)=g(x-X),
\end{equation}
and
\begin{equation}
F(x,X)=g(x-X)+x g'(x-X)
\end{equation}
proves useful.

We now have that
\begin{equation}
\dot p=
- V'(x)+
\int dX\sum_k
c_k F(x,X) Q_k(X,t),
\end{equation}
for the system (we have set $L=1$ for convenience),
and
\begin{equation}
m_k\ddot q_k(X)+m_k\omega_k^2 q_k(X)=c_k x G(x,X),
\end{equation}
for the bath oscillators at location $X$.

The solution for the bath is
\begin{equation}
q_k(t)=
q_k^{(0)}(t)
+
\frac{c_k}{m_k\omega_k}
\int_0^t ds
\sin[\omega_k(t-s)]x(s)G(x(s),X).
\end{equation}
By substituting it back on the equation for the system, we get that
\begin{equation}
\dot p=-V'(x)+F^{II}_0(t)+F^{II}_{fb}(t)+F^{II}_{CT},
\end{equation}
where
\beq
F^{II}_0(t)=
\int dX\sum_k
c_kF(x,X)q_k^{(0)},
\eeq
\beq
F^{II}_{fb}(t) = \int dX\sum_k
\frac{c_k^2}{m_k\omega_k}
F(x,X)
\int_0^t ds
\sin[\omega_k(t-s)]x(s)G(x(s),X),
\eeq
and
\beq
F^{II}_{CT}=
-\int dX \sum_k \frac{c_k^2}{m_k \omega_k^2}x G(x,X) F(x,X).
\eeq

By using Eq.(\ref{eqtruqueseno}), we get
\beq
F^{II}_{fb}(t) = -F^{II}_{CT}+F^{II}_{mem},
\eeq
where
\begin{align}
F^{II}_{mem} &=
-\int dX \sum_k 
\frac{c_k^2}{m_k \omega_k^2}
F
\frac{d}{dt} \int_0^t ds \cos[\omega_k(t-s)] x(s) G(x(s),X) \nn\\
&=-\int dX F(x(t),X) \frac{d}{dt}\int_0^t ds K(t-s) G(x(s),X) x(s).
\end{align}
Again, we have made use of the definition of $K(\tau)$ from Eq.(\ref{defK}).
In the ohmic case, we find
\beq
F^{II}_{mem} = F^{II}_{diss} + \tilde{F}^{II}_0,
\eeq
where
\begin{equation}
    F^{II}_{diss}
    =
    - \eta \int dX F^2(x(t),X) 
    \
    \dot{x}(t),
    \end{equation}
and
    \begin{equation}
   \tilde{F}^{II}_0
   =
    - 2\eta \int dX F(x(0),X)
    G(x(0),X) x(0) \delta(t).
\end{equation}

Finally, we obtain that
\begin{equation}
    M\ddot{x} = -V'(x)+F_{diss}^{II}+F_L^{II}(x,t),
\end{equation}
where 
\begin{equation}
F_{diss}^{II} = - \eta^{II}_{eff}[x] \ \dot{x}, 
\end{equation}
with
\beq
\eta^{II}_{eff}[x] \equiv \eta \int dX F^2(x,X).
\label{eta2eff}
\eeq
By reinserting back $L$, and assuming for instance that
$g(x-X) = \exp[-(x-X)^2/2\sigma^2]$,
we obtain that
$\eta^{II}_{eff}[x] \approx \eta \sqrt{\pi} \sigma/L$
in the 
$\sigma/L \ll 1$
limit.

Following the discussion in Sec.\ref{sec:2.1}, the signature of thermophoresis in the present model is given by the space dependence of 
\begin{align}
F_L^{II}(x,t) &= {F}^{II}_0+\tilde{F}^{II}_0 \nn\\
&=\int dX \sum c_k F(x,X) [Q_k(X,0) \cos(\omega_k t) + (p_k(0)/m_k\omega_k) \sin(\omega_k t)].
\end{align}
Here,
$Q_k(X,0) = q_k(X,0)-\frac{c_k}{m_k\omega_k^2} x(0) g(x(0)-X)$.
Assuming a continuum of thermal baths, each at local equilibrium, amounts to taking
$\langle Q_k(X,0) \rangle = 0 = \langle p_k(X,0) \rangle$,
and
\beq
\langle Q_k(X,0)Q_{k'}(X',0) \rangle = \frac{k_BT(X)}{m_k \omega_k^2}\delta_{kk'}\delta(X-X'),
\eeq
along with
\beq
\langle p_k(X,0)p_{k'}(X',0)\rangle=m_kk_BT(X)\delta_{kk'}\delta(X-X').
\eeq
From that, we get
\begin{align}
\langle F^{II}_L(x,t)F^{II}_L(x',t') \rangle
&=
\int dX \sum_k
\frac{c_k^2}{m_k\omega_k^2}
F(x,X)F(x',X)
k_B T(X)
\cos[\omega_k(t-t')] \nn \\
&=
\int dX 
K(t-t')
F(x(t),X)F(x(t'),X)
k_B T(X).
\end{align}
In the ohmic regime, $K(t-t')=2\eta\delta(t-t')$, so that
\begin{align}
\langle F^{II}_L(x,t)F^{II}_L(x,t') \rangle
&=
2\eta k_B \int dX F^2(x,X) T(X) \delta(t-t')\nn\\
&\equiv 2 D^{II}_{eff}(x) \delta(t-t').
\label{eqdefdeff}
\end{align}
In this model, the diffusion coefficient, 
$D_{eff}^{II}(x)$, 
depends on the temperature field, $T(X)$. 
On the other hand, the effective friction coefficient, 
$\eta_{eff}^{II}[x]$, 
depends on space, but not on the temperature.
In order to better evidence thermophoresis, we derive the Fokker-Planck equation for gCLm-II below.

\subsection{The Fokker-Planck equation for the gCLm-II}

Using Eq.(\ref{eta2eff}), the overdamped regime within the present model reads
\beq
\dot{x} = -\frac{V'(x)}{\eta^{II}_{eff}(x)}+
\sqrt{2D(x)} \xi(t),
\eeq
where $D(x)=D^{II}_{eff}(x)/[\eta^{II}_{eff}(x)]^2$,
and
$\langle \xi(t) \xi(t')\rangle = \delta(t-t')$.
The Fokker-Planck equation in this case is given by
\begin{equation}
\frac{\partial P_{II}(x,t)}{\partial t}
=
\frac{\partial}{\partial x}
\left[
\frac{V'(x)}{\eta^{II}_{eff}(x)}P_{II}(x,t)
+
\frac{\partial}{\partial x}
\left(
D(x)P_{II}(x,t)
\right)
\right].
\end{equation}

If $F(x,X)$ is negligible outside the range $X\approx x$, we can assume a local approximation for the temperature field,
$T(X) \approx T(x)$ 
in 
$D^{II}_{eff}(x)$ 
(see Eq.(\ref{eqdefdeff})).
Otherwise, the diffusion coefficient will depend on the entire vicinity of $x$ determining its effective volume.
In such a local approximation, $T(X) \approx T(x)$, we get that
$D^{II}_{eff}(x) = \eta^{II}_{eff}[x] k_B T(x)$.
Hence,
$D(x) = k_B T(x)/\eta^{II}_{eff}[x]$.
Now,
\beq
\partial_x D(x) = 
\frac{k_B}{\eta^{II}_{eff}}T'(x)
-
\frac{k_B T(t)}{[\eta^{II}_{eff}]^2}
\partial_x\eta^{II}_{eff}.
\eeq
If we can neglect the spatial variation of the effective friction coefficient,
$\partial_x \eta^{II}_{eff}[x] \approx 0$, 
we obtain
\beq
\partial_x P_{ss,II}(x) = - [T'(x)/T(x)] P_{ss,II}(x)
\eeq
as the steady state for a free particle, $V'(x)=0$, as in Eq.(\ref{eqsoret}).
This clearly evidences thermophoresis in the present model.
For instance, let us assume that 
$T(x) = T_0 \exp(-x/L)$, 
which implies that 
$T'/T = -1/L$.
In that case, we find that
\beq
P_{ss,II}(x) \propto \exp(x/L),
\eeq
which means higher concentrations at lower temperature regions.
If instead we cannot neglect 
$\partial_x \eta^{II}_{eff}[x]$, 
we get a more complex scenario, where both $T(X)$ and $\eta^{II}_{eff}[x]$ contribute to the Brownian dynamics.

\section{Conclusions} \label{sec:conclusions}
In summary, we have devised two types of generalized Caldeira-Leggett models that take nonequilibrium baths into account.
Most importantly, we have found signatures of thermophoresis from both models.

Our first model, namely gCLm-I, considers a single bath of oscillators driven out of equilibrium by an external force which, in turn, depends itself on the derivative of the temperature at the system's position.
We find a signature of thermophoresis given by a finite average force acting on the system, $F_{th} = -\kappa T'(x)$.
The diffusion constant does not depend on space, $D_0 = \eta k_B T_0$, thus restricting the model to very small temperature variations.
For the case of a constant thermal gradient, the model admits a Hamiltonian (or Lagrangian) description, a necessary step for the quantum formulation.
In the case of a constant temperature, one recovers the standard Brownian motion.

Our second model, namely gCLm-II, considers a continuous field of baths. 
That is, in each fixed point $X$ in space there exists a thermal bath locally in   equilibrium, at a certain temperature $T(X)$.
A generic function 
$g(x-X)$
sets the spatial distribution of the system-environment coupling, hence establishing an effective volume for the Brownian particle.
Similarly to the standard Caldeira-Leggett model, the local temperature 
$T(X)$ 
is introduced in the statistics of the initial states of the oscillators.
From gCLm-II, we find a space-dependent diffusion coefficient that depends on the local temperatures in the vicinity of the Brownian particle, 
$D_{eff}^{II}(x) = k_B\eta\int dX F^2(x,X)T(X)$.
The signature of thermophoresis is further evidenced from the Fokker-Planck equation corresponding to the gCLm-II.
At the cost of choosing a specific form for the system-baths coupling function, gCLm-II allows for a quantum formulation for a generic temperature field, thus going beyond gCLm-I.
It is still an open question whether our two models can be considered equivalent under certain conditions.
At constant temperatures, we also recover the standard Brownian motion, provided an appropriate choice for 
$g(x-X)$ 
(for instance, a constant function 
$g(x-X)=g_0$ 
would simply reescale the friction coefficient as $\eta_{eff}^{II} = \eta g_0^2$).

In future studies, we aim at addressing the quantum dynamics of both models, as well as other generalized models of the same kind.
This would allow us to describe thermophoresis in quantum Brownian particles, where a continuous distribution of temperature in space might exist, a regime not covered by the quantum master equations employed so far.
Potential applications of the thermophoretic quantum Brownian motion include experimentally relevant setups, such as vortices in Bose-Einstein condensates \cite{castilho}, and quantum solitons undergoing thermophoresis \cite{kim}.

\section*{Acknowledgements} 
\label{sec:acknowledgements}
D.V. was supported by CNPq (Grants 402074/2023-8; 408990/2025-2).
M.M. was supported by CNPq.


%

%

%

%

%
\end{document}